\documentclass[conference, 10pt]{IEEEtran}
\usepackage[compress]{cite}

\usepackage{subfigure}
\usepackage{float}

\usepackage{amsmath,amssymb,amsfonts}
\allowdisplaybreaks
\usepackage{bm}                         
\usepackage{multirow}
\usepackage{cases}                       

\usepackage[normalem]{ulem}             

\newtheorem{lemma}{Lemma}
\newtheorem{theorem}{Theorem}

\newtheorem{remark}{Remark}

\usepackage{algorithmic}
\usepackage{graphicx}
\usepackage{textcomp}
\usepackage{xcolor}
\def\BibTeX{{\rm B\kern-.05em{\sc i\kern-.025em b}\kern-.08em
    T\kern-.1667em\lower.7ex\hbox{E}\kern-.125emX}}

\usepackage{balance}

\begin{document}

\title{Resource Dimensioning for Single-Cell\\Edge Video Analytics
\thanks{
}
}


\author{\IEEEauthorblockN{Jaume Anguera Peris and Viktoria Fodor} 
\IEEEauthorblockA{School of Electrical Engineering and Computer Science\\
KTH Royal Institute of Technology, Stockholm, Sweden\\
Email: \{jaumeap,vfodor\}@kth.se}
}

\maketitle

\begin{abstract}


Edge intelligence is an emerging technology where the base stations located at the edge of the network are equipped with computing units that provide machine learning services to the end users. To provide high-quality services in a cost-efficient way, the wireless and computing resources need to be dimensioned carefully.
In this paper, we address the problem of resource dimensioning in a single-cell system that supports edge video analytics under latency and accuracy constraints. We show that the resource-dimensioning problem can be transformed into a convex optimization problem, and we provide numerical results that give insights into the trade-offs between the wireless and computing resources for varying cell sizes and for varying intensity of incoming tasks. 
Overall, we observe that the wireless and computing resources exhibit opposite trends; the wireless resources favor from smaller cells, where high attenuation losses are avoided, and the computing resources favor from larger cells, where statistical multiplexing allows for computing more tasks. We also show that small cells with low loads have high per-request costs, even when the wireless resources are increased to compensate for the low multiplexing gain at the servers.

\end{abstract}


\section{Introduction}
\label{sec:introduction}
The popularity of mobile devices and the advancements in deep learning algorithms have created ample opportunities for developing applications that extract features and information from images and assist the mobile users in making decisions or obtaining relevant information from their environment. Yet, as these applications become more computationally intensive and delay constrained, it becomes increasingly apparent that mobile devices suffer from insufficient battery capacity, memory, or computational power.


One prominent solution to meet the demands of such delay-sensitive and computationally-complex tasks is to offload them to a server located at the edge of the network. This solution has gathered increasing interest for edge intelligence, which deploys AI-based models at the edge server to process the incoming tasks. The mobile users may then benefit from edge intelligence if the gains from computing at the edge server are greater than the costs of transmitting the information over the wireless link. In recent years, there have been remarkable progress in this field, including techniques for selecting edge servers for fast response times \cite{hamadi2022hybrid}, finding and transmitting only the most informative parts of the image \cite{chen2022context, sun2022elasticedge}, adapting the image format to the network conditions \cite{sun2022elasticedge}, or partitioning the deep learning models to distribute the workload of the image-processing tasks between the mobile users and the edge server \cite{khan2022distributed}.
Despite these advancements, there are still challenges that remain unsolved, such as how to dimension the wireless and the computing resources in the system to ensure that the edge server provides good service quality under acceptable costs. 
This is specially relevant because managing the bandwidth resources at the wireless network and properly allocating the computational resources at the edge server have been shown to be crucial in edge computing and video edge analytics, particularly when the wireless and computing resources may be limited \cite{yi2017lavea,sun2022edgeeye}.

To address the resource-management problem, the literature offers a high variety of solutions. In \cite{liu2018edge}, the authors formulated the trade-off between network latency, computational latency, and accuracy in an edge video-analytic system and proposed a multi-objective optimization problem to select the optimal edge server and video frame resolution. This work was later extended in \cite{wang2020joint} to address the dynamic decisions in the system using queuing theory and to account for the energy consumption of the entire offloading process. More recently, the work in \cite{zhao2022edgeadaptor} has further analyzed the optimal resource provisioning at the edge server when it accommodates a variety of applications, models, and configurations. With this last work, the focus has shifted from proposing the optimal image resolution, offloading frame rate, image encoding, or deep learning algorithm under given wireless and computing resources to understanding the trade-off between the required wireless and computing resources for a given set of user requirements. In this regard, we aim to further extend this line of research and examine the interrelationship between the wireless and computing resources as the system scales in terms of cell size and traffic intensity. 

This paper addresses the resource-dimensioning problem for a single-cell system supporting video analytics. Specifically, we are interested in the minimum amount of wireless and computing resources that need to be deployed in the system to ensure acceptable service quality. For that, we model the entire offloading process under accuracy and latency constraints, and propose an optimization problem to find the relationship between the optimal parameters of the uplink transmission, the edge server, and the deep learning algorithm. We evaluate the optimal solution via numerical results and discuss the trade-off between the different parameters of the system for a cost-efficient resource dimensioning. Overall, our discussion and results provide a clear understanding on how to provision the wireless network and the edge server to meet the demands of video-analytic users, specially when the system scales in size and accommodates a larger amount of incoming tasks.

The remainder of this paper is organized as follows. We describe the system and introduce the mathematical model for the entire offloading process in Sections \ref{sec:system_description} and \ref{sec:system_model}, respectively. Section \ref{sec:optimization_problem} presents the resource-dimensioning problem. Then, we provide the numerical results in Section \ref{sec:numerical_results}, followed by the concluding remarks in Section \ref{sec:conclusions}.


\section{System description}
\label{sec:system_description}
\begin{figure}[t]
    \centering
    \includegraphics[width=0.4\textwidth]{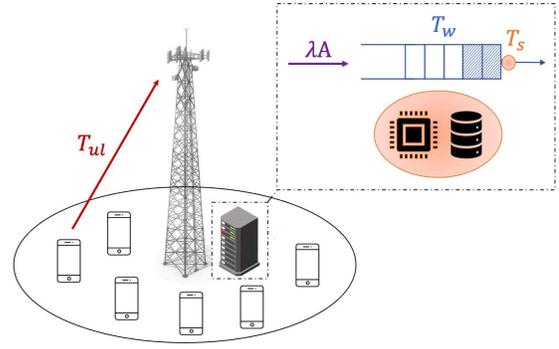}
    \caption{End-to-end system model for a single-cell edge video analytics.}
    \label{fig:systemModel}
\end{figure}

Consider a single-cell cellular system with several mobile users uniformly distributed over the coverage area of the base station, as depicted in Figure \ref{fig:systemModel}. The base station is equipped with a mobile edge computing (MEC) server supporting AI applications for video analytics. These video-analytic applications run an object detection algorithm to assist the edge server and the mobile users in making informed decisions or obtaining relevant information from their environment. With that purpose, mobile users generate image processing tasks and offload them to the MEC server.


Since the amount of data transmitted from the users to the base station is considered to be much larger than the amount of data transmitted from the base station to the users, we focus on the uplink transmission and disregard the little time of eventual feedback to the user. With that in mind, the entire offloading process to perform the video analytics at the MEC server can be divided into two parts. First, users transmit their frame to the base station, and second, the server schedules and processes the frames following a queuing system.

For the uplink transmission, we consider a simple resource allocation where all mobile users receive the same bandwidth $B$ for transmitting a frame to the base station. Moreover, we assume that all video-analytic users can always access the wireless network immediately when needed. This is possible because video-analytic users represent one part of the user population with specific needs and demands, so it is feasible to have a dedicated dynamic network slice that provides them access to the network with high priority \cite{rost2017network}.

For the processing policy at the MEC server, we consider that the edge server is directly connected to the base station via a high-speed link, such that all frames that are completely received at the base station are delivered to the server without delay. The frames are then queued to an infinite buffer and processed on a first-come-first-served basis by a GPU-enabled server with computing capacity $H$.


Altogether, we denote the time to transmit a frame as $T_{ul}$, the time spent in the queue as $T_{w}$, and the time to process a frame as $T_{s}$. We refer to the sum of these three times as the delay of the entire offloading process, and let each user specify a maximum value $D$ for that delay, and a minimum probability $\omega_{\min}$ of satisfying that delay requirement, i.e.,
\begin{equation}
    \mathcal{P}\Big(T_{ul}+T_{w}+T_{s}\leq D\Big) \, \geq \,\omega_{\min}.
    \label{eq:probSuccessfulComputation}
\end{equation}
For simplicity, we consider all users in the network to have the same delay requirement $D$ and $\omega_{\min}$ for all the frames and the same payload for all the image processing tasks. Moreover, we consider that users want the accuracy of the object detection algorithm to be above a minimum threshold $a_{min}$.

Overall, the objective of the resource-dimensioning problem is to find the minimum frequency resource $B$ that should be allocated per frame and the minimum computational resources $H$ that should be placed at the edge server to satisfy the latency and the accuracy requirements for all users and frames in the network. In this regard, the next section presents a comprehensive system model and derives the expressions of the parameters that characterize both the uplink transmission and the MEC server.

\section{System model}
\label{sec:system_model}
\subsection{Uplink communication and transmission time}
For the uplink transmission, we consider a free-space propagation model with frequency-division multiple access and i.i.d.\@ Rayleigh fading. All mobile users utilize a maximum-ratio transmission precoding vector, such that the effect of the channel can be modelled at the receiver by the multiplicative Exponential random variable $|g|^2\sim \exp{(\gamma)}$, where $\gamma=\frac{\lambda_c^2}{16\pi^2}$, and $\lambda_c$ is the wavelength of the carrier frequency. Moreover, we consider that all users utilize distance-proportional fractional power control of the form $P r^{\alpha\epsilon}$, where $P$ is the reference power at $1$ kilometer from the base station, $r$ is the distance to the base station, $\alpha$ is the path-loss exponent, and $\epsilon\in[0,1]$ is the power control coefficient. To model the maximum transmission power of the mobile users, we consider the transmit powers $Pr^{\alpha\epsilon}$ to be limited by the peak power $\Bar{P}$.
Finally, we consider the noise power at the receiver to be $\sigma^2$. Under this transmission model, the signal-to-noise ratio (SNR) at the base station for any randomly selected user is
\begin{equation}
    \text{SNR} = \frac{|g|^2 \,\ell(r,\alpha,\epsilon)\,r^{-\alpha}}{\sigma^2},
    \label{eq:SNR}
\end{equation}
where 
\begin{equation}
    \ell(r,\alpha,\epsilon) = \min\left(Pr^{\alpha\epsilon},\Bar{P}\right).
    \label{eq:peakPowerConstrain}
\end{equation}

From there, we consider that any active user with given bandwidth $B$ estimates the state of the channel from pilot signals and adjusts its modulation and coding according to its instantaneous SNR. As a result, any active user achieves a transmission rate over a single coherence block close to the Shannon capacity, $B\log_2(1+\text{SNR})$. At the same time, video analytics are typically optimized to operate with frame sizes on the order of hundreds of thousands of pixels \cite{ibm2021frameresolution}. Hence, the frames are sufficiently large for the transmission to take place over several coherence blocks, such that the transmission signals experience a high variety of fading realizations and the users experience the ergodic capacity of the channel.
With that in mind, let us now calculate the ergodic capacity for any user transmitting a frame at distance $r$ from the base station as \cite[Theorem 3]{andrews2011tractable}
\begin{align}
    \Bar{R} & = \mathbb{E}[B \log_2(1+\text{SNR})] \nonumber \\
    & \overset{(a)}{=} \frac{B}{\log(2)} \int_{0}^{\infty} \mathcal{P}\left(\log(1+\text{SNR})\geq t\right)dt \nonumber \\
    & \overset{(b)}{=} \frac{B}{\log(2)} \int_{0}^{\infty} \mathcal{P}\left( |g|^2 \geq \frac{(e^t-1)\, \sigma^2}{\ell(r,\alpha,\epsilon)\,r^{-\alpha}} \right)dt \nonumber \\
    & \overset{(c)}{=} \frac{B}{\log(2)} \int_{0}^{\infty} \exp\left(- \frac{(e^t-1)\, r^{\alpha}}{\gamma\,\ell(r,\alpha,\epsilon)}\sigma^2\right) dt \nonumber \\
    & \overset{(d)}{=} \frac{B}{\log(2)} \, \exp\left(\frac{B N_0\,r^{\alpha}}{\gamma\,\ell(r,\alpha,\epsilon)}\right) \, E_1\left(\frac{B N_0\,r^{\alpha}}{\gamma\,\ell(r,\alpha,\epsilon)}\right),
    \label{eq:ergodicCapacity}
\end{align}
where $(a)$ follows from the fact that the expectation of any positive random variable $X$ can be calculated in the sense of Lebesgue-Stieltjes as $\mathbb{E}[X]=\int_0^\infty \mathcal{P}\left(X\geq t\right)dt$, $(b)$ follows from the definition of the SNR in \eqref{eq:SNR}, $(c)$ follows from the exponential distribution of $|g|^2$, and $(d)$ follows from the definition of the exponential integral function $E_1(x) = \int_x^\infty \frac{e^{-t}}{t}\, dt$. Notice that we expressed the noise power $\sigma^2$ in \eqref{eq:ergodicCapacity} as the product between the noise power spectral density, $N_0$, and the allocated bandwidth $B$.

Now that we have characterized the uplink communication, let us proceed with the uplink transmission time. To generate the image processing task, users offload frames with resolution $s\times s$ pixels, encoded at $\theta$ bits per pixel, and compressed at rate $\xi$:1, i.e., frames represented by $\theta s^2/\xi$ bits. From there, and the assumption that users experience the ergodic nature of the channel, it follows from the analysis in \cite{peris2022modelling} that the probability density function of the uplink transmission time over the entire uplink transmission for any user at distance $r$ from the base station can be approximated as
\begin{equation}
    f_{T_{ul}}(t) \longrightarrow \Delta\left(t-\frac{\theta s^2}{\xi\,\Bar{R}}\right),
    \label{eq:PDF_uplink}
\end{equation}
where $\Delta(\cdot)$ represents the Dirac delta function.

\subsection{Edge analytics and service time}
To perform the video analytics, the edge server utilizes a convolutional neural network pre-trained with YOLOv5 \cite{glenn2021yolov5}. Since YOLOv5 is capable of handling frames of different resolutions without changing its associated learnable parameters, it is possible to parametrize the inference time of the detection algorithm \cite{liu2018edge}, such that we can express the service time as
\begin{equation}
    T_{s} = \frac{c_1s^3 + c_2}{H}
    \label{eq:serviceTime}
\end{equation}
for some positive constants $c_1$, $c_2$, where the units of the numerator are number of floating-point operations (in trillions).

Similarly, it is possible to parameterize the accuracy of the object detection algorithm as
\begin{equation}
    a(s) = c_3 - c_4e^{-c_5 s},
    \label{eq:accuracy_YOLO}
\end{equation}
for some other positive constants $c_3$, $c_4$, and $c_5$, where the accuracy of the detection is measured as the mean average precision of the object detection algorithm for a predefined threshold of the intersection over union \cite{liu2018edge}. For more information about the effect of the learning and inference processes on the constants $c_1,\dots,c_5$, please refer to \cite{bochkovskiy2020yolov4, ameur2021deployability}.

\subsection{Service policy and waiting time}
\label{sec:policy-Waiting}

We consider that users operate independently of each other and are uniformly distributed over a circular cell with area $A$. The traffic generated per unit area is Poisson distributed with intensity $\lambda\, \left[\text{frames}\cdot s^{-1}\cdot km^{-2}\right]$, and frames arrive at the MEC servers after a random $T_{ul}$ time. Since the Poisson process is homogeneous in time, the process remains translation invariant in time \cite{baccelli2010stochastic}, and the arrival process at the edge server is Poisson distributed with intensity $\lambda A$.

Upon arrival, the frames are delivered on an infinite buffer and then wait to be processed on a first-come-first-served basis. The processing time $T_s$ is deterministic and is determined by the resolution of the transmitted frames, the neural network considered for the video analytics, and the available computational resources at the edge server, as shown in \eqref{eq:serviceTime}. As a result, it yields that server can be modelled as an M/D/1 queuing system with load $\rho = \lambda A\, T_{s}$.


Considering the M/D/1 queuing system, the cumulative density function of $T_w$ can be derived from the state probabilities using the Erlang’s principle of statistical equilibrium \cite[Sections 10.4.2 and 10.4.4]{iversen2010teletraffic}. Specially, for a given load $\rho$ and service time $T_s$, the complementary cumulative distribution function (CCDF) of the waiting time can be expressed as
\begin{equation}
    \mathcal{P}(T_w>T)=1-(1-\rho)\sum_{\nu=0}^{\Tilde{T}}\frac{\left[\rho(\nu-t)\right]^\nu}{\nu!}\,e^{-\rho(\nu-t)},
    \label{eq:PDF_wait}
\end{equation}
for any $T\geq 0$, where $t = \frac{T}{T_s}$, $\Tilde{T}=\left\lfloor t \right\rfloor$, and $\left\lfloor \cdot \right\rfloor$ is the greatest integer function. In terms of the function's behaviour, \eqref{eq:PDF_wait} is continuous and monotonically decreasing for any $\rho$ and $T_s$, and is concave for $T\in [0,T_s)$ and convex for $T\in [T_s,\infty)$.

Note, however, that the concavity of $\mathcal{P}(T_w> T)$ poses a hindrance to formulating the optimization problem as a convex problem. Hence, we consider instead a convex approximation for \eqref{eq:PDF_wait}. Among all possible approximations, we select Henk's approximation \cite{tijms2003first}, as it offers a tighter approximation to the true CCDF for higher $\rho$ and for any $T$ close to or higher than $T_s$. Considering this, and following the analysis in \cite[Section 9.6.1]{tijms2003first},
we get the following expression for the M/D/1 queue
\begin{equation}
    \mathcal{P}^{\textit{Henk}}(T_{w} > T) = \frac{(1-\rho)}{(\rho\tau-1)}e^{-\lambda A(\tau-1)T},\quad \forall\, T\geq0,
    \label{eq:PDF_wait_HenkApprox}
\end{equation}
where $\tau = -W_{-1}(-\rho e^{-\rho})/\rho$, 
and $W_{-1}(\cdot)$ is the negative branch of the Lambert function.

\section{Resource-dimensioning problem}
\label{sec:optimization_problem}
Now that we have characterized the entire offloading process, we are ready to construct the optimization problem that minimizes the cost of the frequency and the computational resources. In the considered edge-analytics system, each frame receives the same bandwidth resources and all users utilize the same power control coefficient for the uplink transmission, so the users that experience the largest $T_{ul}$ are the users located at the edge of the cell. As the other delays are independent of the users' location, the edge users always experience the worst overall performance. Therefore, we formulate the optimization problem with the constraints that all the frames arriving from these cell-edge users satisfy the latency requirement $D$ with probability  $\omega_{\min}$, and satisfy the minimum object detection accuracy $a_{\min}$. That is to say, if the latency and accuracy requirements of the frames of the cell-edge user are satisfied for some $B$ and $H$, then the latency and accuracy requirements of all the other users and frames will be satisfied as well for the same $B$ and $H$.

With that, let us now formulate the resource-dimensioning problem. First, to ease the notation and facilitate the analysis of the convexity of the optimization problem, we decouple \eqref{eq:probSuccessfulComputation} into two parts by introducing a free variable $T$,
\begin{equation}
    \mathcal{P}(T_{w}>T) \leq 1-\omega_{\min}\quad \textit{and} \quad T_{ul} + T + T_s = D.
    \label{eq:separateConditions}
\end{equation}
Then, we adopt the weighted sum method \cite{marler2010weighted} to accurately represent the complex preferences between the wireless and computing resources and construct a multi-objective function that aims at minimizing the cost of the wireless and computing resources per frame,
\begin{equation}
    J(B,H|\lambda, r, \beta_1, \beta_2) = \beta_1 B + (1-\beta_1)\, \beta_2\, \frac{H}{\lambda \pi r^2}.
    \label{eq:objective_function_def}
\end{equation}
In particular, the wireless resource per frame is $B$, and the computing resource $H$ is shared by the requests from all active users, giving an average computing resource per request $H_f=H/(\lambda \pi r^2)$. The positive weight parameter $\beta_1\in(0,1)$ characterizes the trade-off between $B$ and $H_f$, and the positive parameter $\beta_2$ makes sure that the model has an adequate representation of the objective function by bringing $B$ and $H_f$ to the same order of magnitude.

After putting all together and combining the results from the previous sections, the multi-objective optimization problem for the single-cell system can be expressed for a fixed $r$, $\lambda$, $D$, $\omega_{\min}$, $a_{\min}$, and $\rho_{\max}$ as
\begin{subequations}
    \begin{align}
    \min_{\{B,H,T,s\}} \quad & \beta_1 B + (1-\beta_1)\, \beta_2\, \frac{H}{\lambda \pi r^2} \label{eq:objective_function_original} \\
    \textrm{s.t.} \quad & \mathcal{P}(T_{w} > T) \leq 1 - \omega_{\min}, \label{eq:constrain1_original}\\
                        & \frac{s^2 \kappa_1}{\phi(B,\kappa_2)} + T +  \frac{c_1 s^3 + c_2}{H} = D, \label{eq:constrain2_original}\\
                        & \kappa_3 (c_1 s^3 + c_2) \leq H, \label{eq:constrain3_original}\\
                        & B > 0, \; H > 0, \; T \geq 0,\; s\geq \kappa_4,  \label{eq:constrain4_original}
    \end{align}
    \label{eq:optimization_problem_original}
\end{subequations}
\noindent where
\begin{equation*}
    \phi(B,\kappa_2) = B \exp\left(B\kappa_2\right) E_1\left(B\kappa_2\right),
\end{equation*}
and
\begin{alignat}{2}
    \kappa_1 &= \frac{\theta}{\xi} \log(2), \quad & \kappa_2 &= \frac{N_0\,r^{\alpha}}{\gamma\, \ell(r,\alpha,\epsilon)}, \label{eq:kappa_def}\\
    \kappa_3 &=  \frac{\lambda \pi r^2}{\rho_{\max}}, \quad & \kappa_4 &= \frac{1}{c_5}\ln\left(\frac{c_4}{c_3-a_{\min}}\right) \nonumber
\end{alignat}
are just constants.

Specially, the optimization problem in \eqref{eq:optimization_problem_original} aims at dimensioning the wireless and computing resources of a single-cell edge-analytic system such that the edge users, and consequently all the other users within the cell, have a guaranteed probability $\omega_{\min}$ of successfully offloading and processing their frames within the delay requirement $D$, and a guaranteed accuracy $a_{\min}$ for the AI-based object detection algorithm. The optimization variables are $B$, $H$, $T$, and $s$. Constraints \eqref{eq:constrain1_original} and \eqref{eq:constrain2_original} guarantee that the optimal $B$ and $H$ satisfy the latency requirements for each user and frame, and constraint \eqref{eq:constrain3_original} guarantees that the server satisfies the demands of all users and does not overload. Finally, constraint \eqref{eq:constrain4_original} limits the domain of the variables and guarantees that the optimal $s$ satisfies the accuracy requirements for each user and frame.


As discussed in Section \ref{sec:policy-Waiting}, the cumulative density function of $T_w$ is non-convex over the domain of $T$, so we instead consider the Henk's approximation \eqref{eq:PDF_wait_HenkApprox}. Similarly, the equality \eqref{eq:constrain2_original} introduces a non-affine condition to our optimization problem, which may result in a non-convex feasible region. Hence, we substitute the equality with an inequality. This change does not alter the optimal result in any way, as we are trying to minimize $B$ and $H$, so there is no incentive to converge to a solution where the total delay of the entire offloading process is less than $D$. Finally, to preserve the convexity of $\phi(B,\kappa_2)$, we utilize the epigraph representation of the $\min$ operator of the peak power constrain \eqref{eq:peakPowerConstrain}.

Considering these, we can re-formulate the optimization problem for a fixed $r$, $\lambda$, $D$, $\omega_{\min}$, $a_{\min}$, and $\rho_{\max}$ as
\begin{subequations}
    \begin{align}
    \min_{\{B,H,T,s\}} \quad & \beta_1 B + (1-\beta_1)\, \beta_2\, \frac{H}{\lambda \pi r^2} \label{eq:objective_function} \\
    \textrm{s.t.} \quad & \mathcal{P}^{\textit{Henk}}(T_{w} > T) \leq 1 - \omega_{\min}, \label{eq:constrain1}\\
                        & \frac{s^2 \kappa_1}{\phi_{\textit{low}}(B,\kappa_2)} + T + \frac{c_1 s^3 + c_2}{H} \leq D \label{eq:constrain2}\\
                        & \frac{s^2 \kappa_1}{\phi_{\textit{peak}}(B,\kappa_2)} + T + \frac{c_1 s^3 + c_2}{H} \leq D \label{eq:constrain3}\\
                        & \eqref{eq:constrain3_original}-\eqref{eq:constrain4_original}\nonumber.
    \end{align}
    \label{eq:optimization_problem}
\end{subequations}
where
\begin{align*}
    \phi_{\textit{low}}(B,\kappa_2) &= \Big\{ \phi(B,\kappa_2) \,:\, \ell(r,\alpha,\epsilon)=Pr^{\alpha\epsilon}\Big\}, \\
    \phi_{\textit{peak}}(B,\kappa_2) &= \Big\{ \phi(B,\kappa_2) \,:\, \ell(r,\alpha,\epsilon)=\bar{P}\Big\}.
\end{align*}

\begin{lemma}
    \label{lemma:convexAnalysis}
    \textit{The problem formulated in \eqref{eq:optimization_problem} is convex.}
    \begin{IEEEproof}
        The objective function is an affine function, and the feasible region is jointly convex on all variables because all constraints are convex. Therefore, \eqref{eq:optimization_problem} is a convex optimization problem.
    \end{IEEEproof}
\end{lemma}
\begin{remark}
    \label{remark:extension_optProblem}
    \textit{The problem formulation could be further extended to consider three more limitations. First, a maximum bandwidth allocation per frame to account for the limitations of the user equipment. Second, a maximum processing capacity at the server to account for energy or cost limitations. Third, an irregular coverage area to account for fading, shadowing, or asymmetric channel conditions. In either case, since this paper aims to find the pure relationship between $B$ and $H$ and serve as a basis for more complex problems, we leave these possible extensions for future work.}
\end{remark}

Since we are formulating the optimization problem with the Henk's approximation of the CCDF of the waiting time, it is possible that the optimal solution $(B^*, H^*, T^*, s^*)$ for the convex problem \eqref{eq:optimization_problem} is neither a feasible point nor an optimal solution for the optimization problem \eqref{eq:optimization_problem_original}. However, we can calculate and compensate for this approximation error in advance as stated below.


\begin{theorem}
    \label{theorem:errorApprox_pWait}
    \textit{The error incurred from using the Henk's approximation can be calculated in advance. Specially, we can compensate for this approximation error by adjusting the delay requirement to $\omega_{\min}+0.017$ to make sure that the optimal solution for the convex problem \eqref{eq:optimization_problem} is always a feasible solution for the original problem \eqref{eq:optimization_problem_original}}.
    \vspace{2pt}
    \begin{IEEEproof}
    Define the maximum error between the true CCDF of the waiting time and the Henk's approximation as
    \begin{equation}
        e^*(\rho) = \max_T \,\Big( \mathcal{P}(T_w > T) - \mathcal{P}^\textit{Henk}(T_w>T)\Big).
    \end{equation}
    Note that we are not interested in the absolute value of the difference because the only cases for which the optimal solution of the convex problem does not satisfy constrain \eqref{eq:constrain1} are those where the Henk's approximation takes lower values than the true CCDF for the same $T$. Also, note that the maximum error does not depend on the service time $T_s$.
    
    Considering this, we can take the first derivative and find that the argument that maximizes the error is
    \begin{equation*}
        T_{\textit{maxError}} = -\frac{T_s}{\rho \tau}\log\left(\frac{\rho\tau-1}{\tau-1}\right).
    \end{equation*}
    Then, if we numerically evaluate $e^*(\rho)$ at $T_{\textit{maxError}}$, we can find that $e^*(\rho) \leq 0.017,\;\forall\rho $. Hence, the maximum difference between the true CCDF of the waiting time and the Henk's approximation is at most $0.017$ for any server load and service time. Consequently, if we set the new constrain \eqref{eq:constrain1} to be $\omega_{\min}+0.017$, the optimal solution for the convex problem \eqref{eq:optimization_problem} will always be a feasible solution for the original problem \eqref{eq:optimization_problem_original}. This concludes the proof.
    \end{IEEEproof}
\end{theorem}

Finally, it is important to mention that the problem formulation \eqref{eq:optimization_problem} does not lead to closed-form solutions of the optimal $B$, $H$, or $H_f$ values. However, to understand the effect of the joint dimensioning of these resources, it is meaningful to evaluate how they need to be scaled when one of them can be considered costless. 

\begin{theorem}
    \label{theorem:minResources}
    \textit{The minimum wireless resources per frame and the minimum computational resource at the MEC server that are required to satisfy the latency and accuracy requirements are given by
    \begin{align*}
        B|_{\min} &= \frac{-\Omega_1/\kappa_2}{\Omega_1 + W_{-1}\left(-\Omega_1\,\exp\left(-\Omega_1\right)\right)},\\[.2cm]
        H|_{\min} & = \Omega_2 \times \max\left\{\kappa_3, \frac{1}{D\left(1-\Omega_1\right)}\right\},
    \end{align*}
    where
    \begin{equation*}
        \Omega_1 = \kappa_2 \frac{\kappa_1\kappa_4^2}{D}, \quad\; \Omega_2 = c_1\kappa_4^3+c_2,
    \end{equation*}
    and $\kappa_1$, $\kappa_2$, $\kappa_3$, and $\kappa_4$ are defined in \eqref{eq:kappa_def}.}
    \begin{IEEEproof}
    The minimum number of frequency resources can be derived by assuming an infinite number of computational resources and by finding an upper bound of the ergodic capacity using Jensen's inequality. Similarly, the minimum number of computational resources can be derived by assuming an infinite number of frequency resources. The expressions then follow from these assumptions and the constraints of the optimization problem given in \eqref{eq:optimization_problem}.
    \end{IEEEproof}
\end{theorem}

\section{Numerical results}
\label{sec:numerical_results}
\begin{table}[t]
    \centering
    \caption{System parameters}
    \begin{tabular}{|l|c|}
        \hline
        Path-loss exponent $\alpha$ & 4 \\ \hline
        Power control coefficient $\epsilon$ & 0.5 \\ \hline
        Peak transmission power $\Bar{P}$ & 23 dBm (200 mW) \\ \hline
        Transmission power per unit distance $P$ & 10 dBm/km (10 mW/km) \\ \hline
        Noise power spectral density $N_0$ & -174 dBm/Hz \\ \hline
        Carrier frequency $f_c$ & 2.4 GHz \\ \hline
        Frame encoding rate $\theta$ & 24 bits/pixel \\ \hline
        Frame compression rate $\xi$ & 2 \\ \hline
        Maximum load at the server $\rho_{\max}$ & $0.99$ Erlangs\\ \hline
        Maximum delay requirement $D$ & $500$ ms\\ \hline
        Minimum probability of success $\omega_{\min}$ & $0.8$ \\ \hline
        Minimum object detection accuracy $a_{\min}$ & $0.9$ \\ \hline
        Trade-off parameter obj. function $\beta_1$ & $0.5$  \\ \hline
        Scaling parameter obj. function $\beta_2$ & $10^6$  \\ \hline
    \end{tabular}
    \label{tab:system_parameters_simulation}
\end{table}

This section evaluates the optimal resource dimensioning of a single-cell system supporting edge video analytics. Specially, we solve the optimization problem in \eqref{eq:optimization_problem} using the convex optimization CVXOPT toolbox from MATLAB. To account for the two mathematical functions that are not supported in CVXOPT, namely, the negative branch of the Lambert Function, $W_{-1}(\cdot)$, and the exponential integral function, $E_1(\cdot)$, we consider their equivalent expressions from \cite[Eq.~5]{barry2000analytical} and \cite[Eq.~5.1.11]{abramowitz1988handbook}, respectively. With that, we analyze the optimal solution of the optimization problem and discuss the effect of different parameters on the design of cost-efficient strategies. In all cases, we consider the system parameters from Table \ref{tab:system_parameters_simulation}, corresponding to a typical cellular network \cite{ghosh2010fundamentals} and a typical video-analytic use-case \cite{ran2018deepdecision}, unless stated otherwise. For the detection algorithm, we set the parameters that define \eqref{eq:serviceTime} and \eqref{eq:accuracy_YOLO} to $c_1 = 7\cdot 10^{-10}$, $c_2=0.083$, $c_3=1$, $c_4=1.578$, and $c_5=6.5\cdot 10^{-3}$, which have been shown to fit the ground truth with a root mean square error less than $0.03$ \cite{liu2018edge}. Also, recall that we assumed for simplicity that all frames have the same maximum delay requirements $D$ and $\omega_{\min}$, and the same minimum object detection accuracy $a_{\min}$.

\begin{figure*}
    \centering
    \subfigure[]{\includegraphics[width=0.45\textwidth]{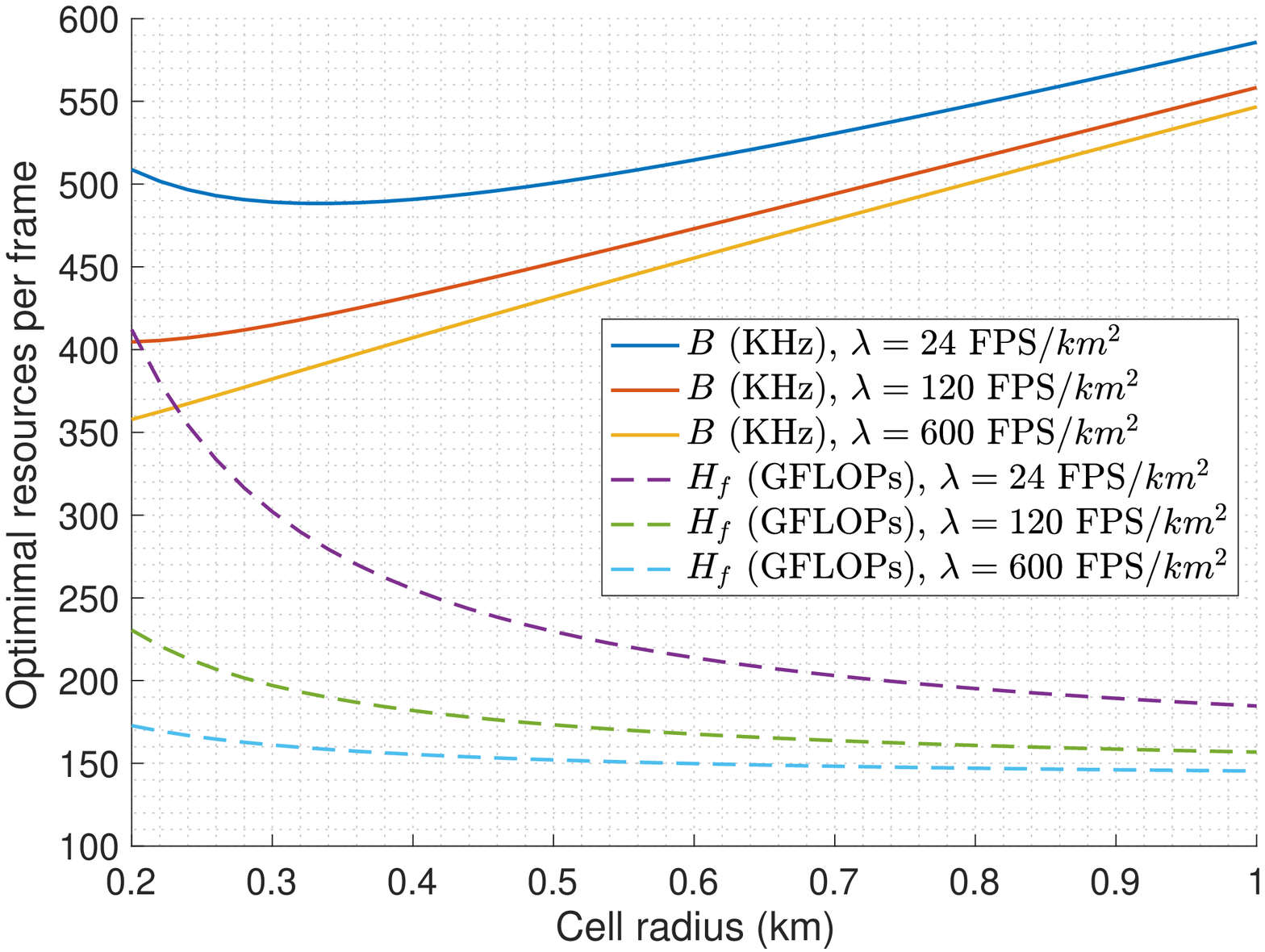} \label{fig:optBH_cellRadius}}\hspace{.4cm}
    \subfigure[]{\includegraphics[width=0.45\textwidth]{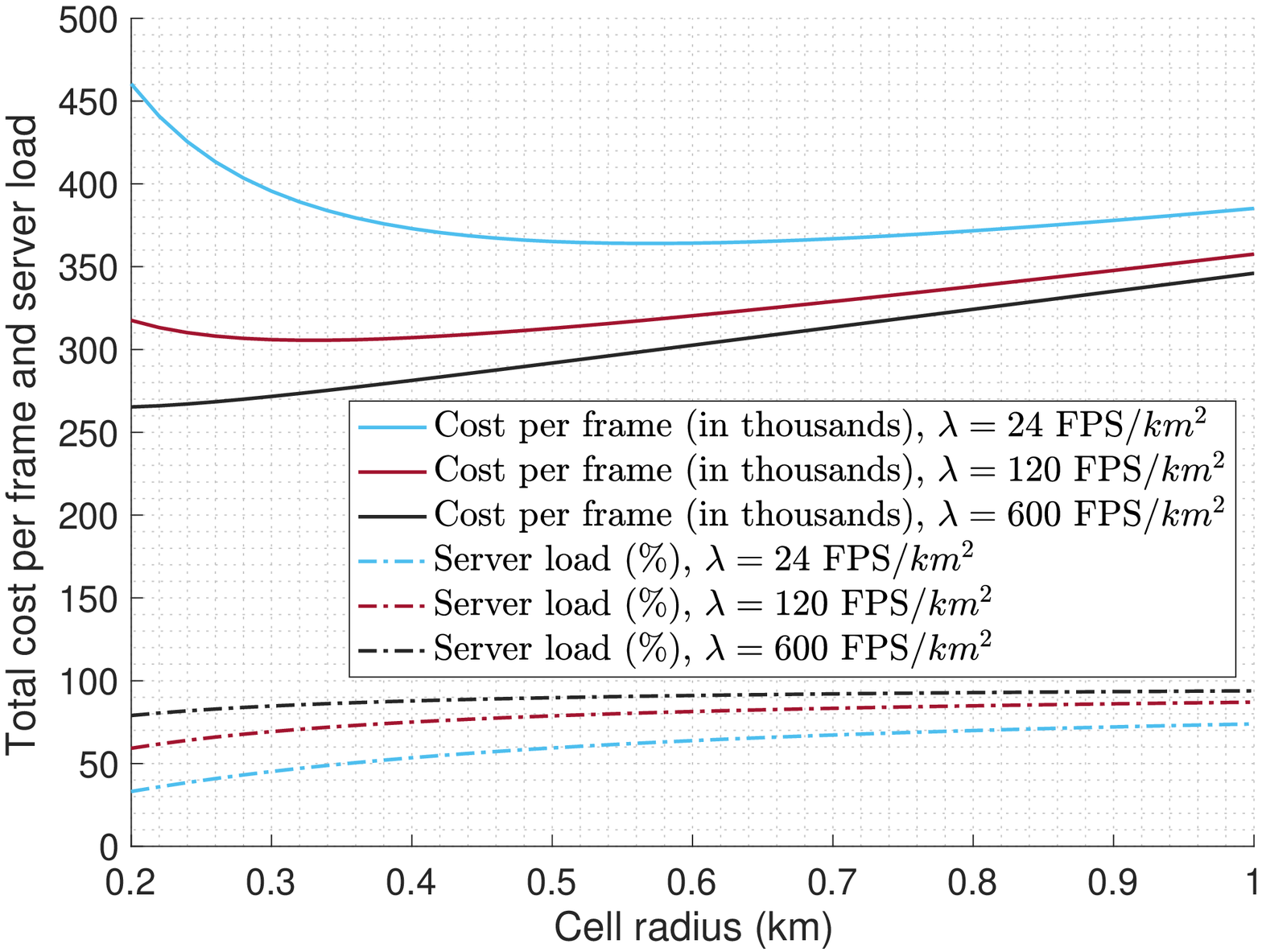} \label{fig:optValueLoad_cellRadius}}
    \subfigure[]{\includegraphics[width=0.45\textwidth]{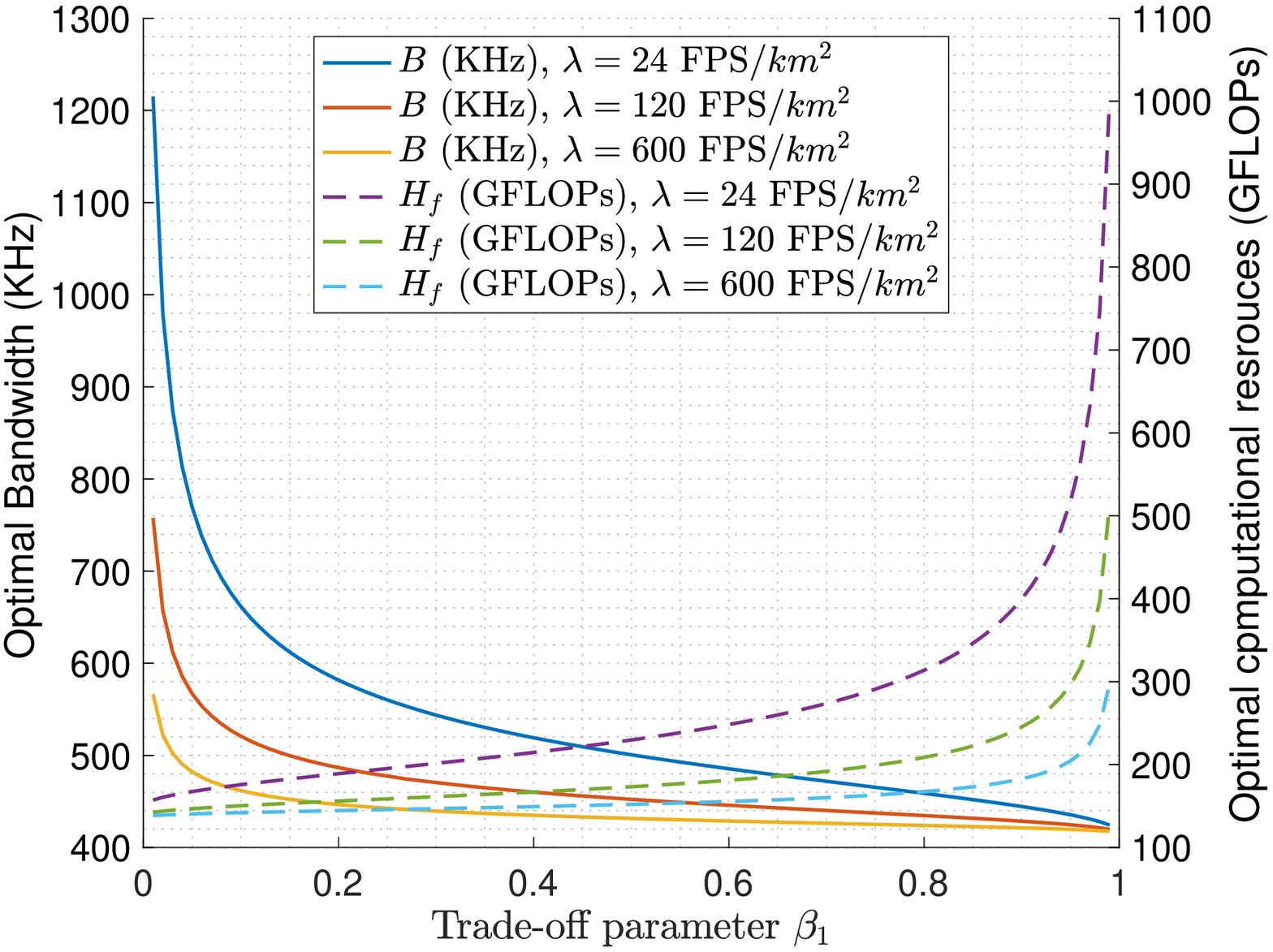} \label{fig:optBH_betaTradeOff}}\hspace{.4cm}
    \subfigure[]{\includegraphics[width=0.45\textwidth]{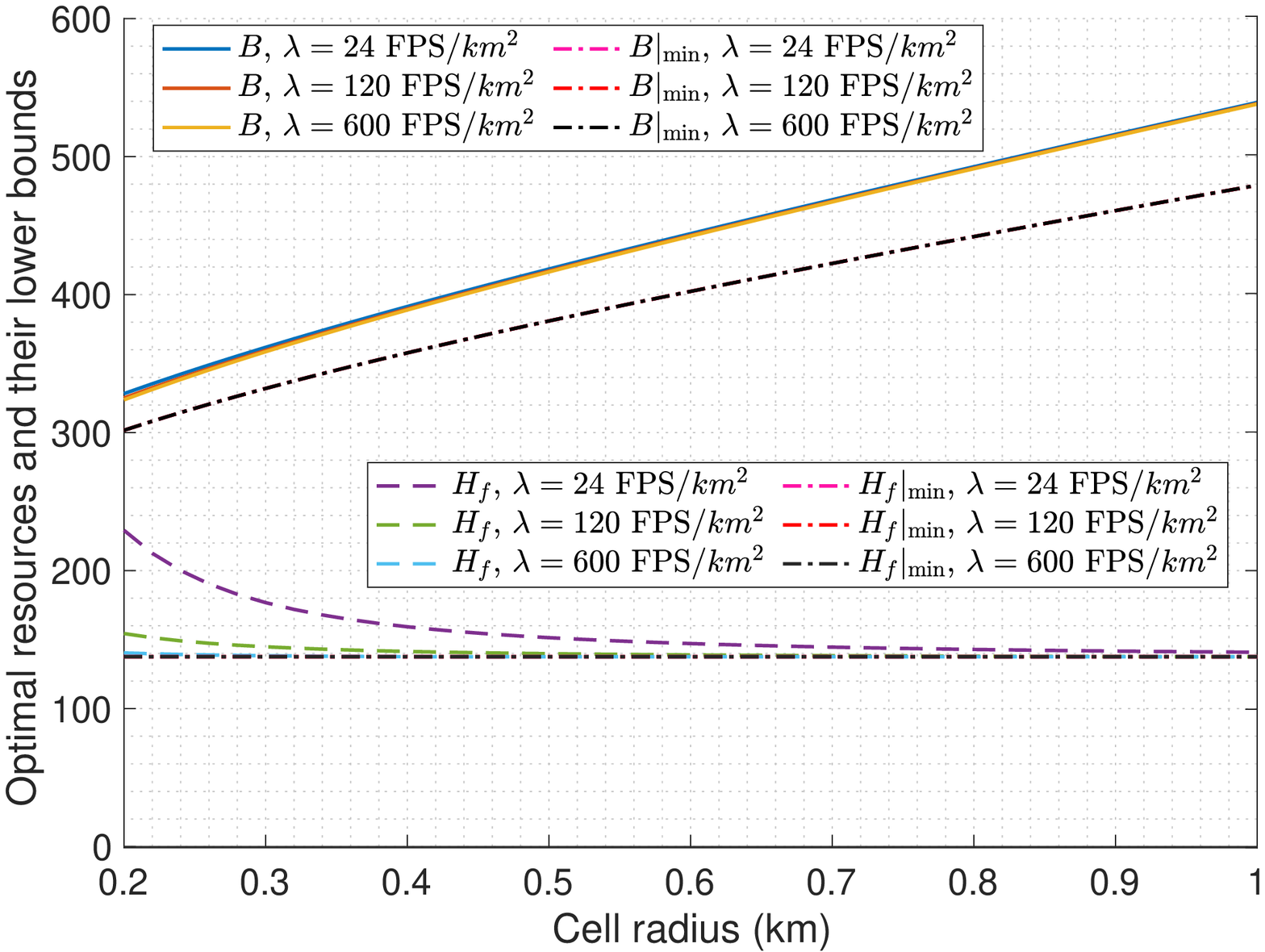} \label{fig:optBHLowerBound_cellRadius}}
    \caption{(a)--(b) Optimal wireless and computing resources per frame, optimal server load, and total cost of the resource-dimensioning (in thousands) as a function of the cell radius $r$ for different traffic intensities $\lambda$. (c) Effect of the trade-off parameter $\beta_1$ on the optimal resources for a single-cell system with radius $r=0.5$ km and for different traffic intensities $\lambda$. (d) Optimal wireless resources per frame (in KHz) for $\beta_1\rightarrow0$, optimal computing resources per server (in GFLOPs) for $\beta_1\rightarrow1$, and their corresponding theoretical values as a function of the cell radius $r$ for different traffic intensities $\lambda$.}
    \label{fig:simulation_results}
\end{figure*}

Figure \ref{fig:optBH_cellRadius} shows the optimal bandwidth and computing resources per frame, namely, $B$ and $H_f = H/(\lambda\pi r^2)$, as a function of the cell radius $r$ for different traffic intensities $\lambda$. If we analyze the resources jointly, we observe a multiplexing effect for both the frequency and the computational resources; the higher the traffic in the system, the lower the number of resources required per frame. If we analyze the resources individually, we observe that the bandwidth resources benefit from having small cell radii due to lower signal attenuation, whereas the computing resources benefit from higher cell radii due to increased multiplexing gain. We see the effect of joint dimensioning when the cell sizes are small and the traffic intensity is low. In this scenario, the server needs a significant amount of resources to make sure that a frame satisfies the delay requirements whenever it arrives to the queue, but that, in turn, results in an underutilized server because the arrivals are less frequent. As the use of shared computational resources is very inefficient in this case, the per-frame bandwidth allocation needs to be increased to avoid very high computation costs. Consequently, the bandwidth needs are not monotonously increasing for increasing $r$.

Figure \ref{fig:optValueLoad_cellRadius} shows the optimal value of the objective function \eqref{eq:objective_function_def} and the optimal server load as a function of the cell radius $r$ for different traffic intensities $\lambda$. The figure demonstrates that the server needs to be over-dimensioned significantly in small cells under low load. We also observe that a single-cell system with moderate or higher traffic behaves quite similarly for increasing $r$, but these are distinct from the low traffic case, where small cells become very expensive, despite the good transmission conditions.


Figure \ref{fig:optBH_betaTradeOff} shows the optimal resource dimensioning for a single-cell system with cell radius $r=0.5$ km as a function of the trade-off parameter $\beta_1$ for different traffic intensities $\lambda$. Overall, smaller $\beta_1$ prioritize having more frequency resources than computational resources and vice-versa. In the extremes, the parameter that has decreasing weight in the objective function grows without bound, and the other parameter becomes the minimum resources per frame that should be allocated to satisfy the delay requirements of all the users. Notice also that the figure shows symmetry around $\beta_1=0.5$, thus corroborating the adequate choice of the scaling parameter $\beta_2$.


Finally, on Figure \ref{fig:optBHLowerBound_cellRadius} we evaluate the optimal bandwidth resource and computing resource per frame in the extreme cases $\beta_1\rightarrow0$ and $\beta_1\rightarrow1$. The optimal values from \eqref{eq:optimization_problem} are shown as a function of the cell radius $r$ for different traffic intensities $\lambda$, and are compared to the theoretical lower bounds from Theorem~\ref{theorem:minResources}, with $H_f|_{\min}=H|_{\min}/(\lambda\pi r^2)$.
%
%
We observe that the bandwidth resources per frame become larger for increasing cell radii and do not depend on the traffic in the system. These results are contrary to the observations from Figure~\ref{fig:optBH_cellRadius}, where the computational costs are also taken into account and the bandwidth resources compensate for the inefficient computation. In addition, we observe that the computing resources per frame exhibit a moderate increase for small cell sizes, and remain nearly constant for increasing traffic intensity $\lambda A$. This is again contrary to the observations from Figure~\ref{fig:optBH_cellRadius}, where the wireless costs are also taken into account and the computing resources are tuned to obtain waiting times that satisfy the minimum success probability.
In all cases, the theoretical equivalents are close to the optimal values obtained from the optimization problem, with $H_f|_{\min}$ being a tighter approximation to $H_f$ than $B|_{\min}$ to $B$. This is because the minimum computing resources are derived directly from the constraints of the problem, whereas the minimum bandwidth resources requires the Jensen's inequality to find a closed-form expression.


\section{Conclusions}
\label{sec:conclusions}
This paper extends a line of research in edge video analytics in which the focus is no longer on adjusting the demands of the users according to the available resources in the system, but rather on dimensioning the wireless and computing resources in the system to meet the demands of the mobile users.
For that, we presented an entire end-to-end system model and proposed a multi-objective optimization problem to characterize the trade-off between the wireless and computing resources
under latency and accuracy constraints. We showed that the resource-dimensioning problem is non-convex, but can be reformulated into a convex problem.
The results show that the wireless and computing resources exhibit opposite trends; the wireless resource favors from smaller cells because of lower signal attenuation, and the computing resources favors from bigger cells because the server avoids being under-utilized. Besides, the wireless and computing resources benefit from higher traffic, as the system can favourably multiplex the incoming tasks. We also derived expressions for the minimum resources needed in the system to meet the demands of the users. We showed that the minimum wireless resources only depend on the distance between the user and the base station, and that the minimum computing resources exhibit a moderate increase for small cell sizes and remain nearly invariant for high traffic arrival intensities.
%
Overall, we conclude that the incoming traffic at the edge server, the cell radius, and the cost of the bandwidth and the computational capacity are decisive when dimensioning the resources of single-cell systems supporting edge video analytics. 

\bibliography{references}

\begin{thebibliography}{10}
\providecommand{\url}[1]{#1}
\csname url@samestyle\endcsname
\providecommand{\newblock}{\relax}
\providecommand{\bibinfo}[2]{#2}
\providecommand{\BIBentrySTDinterwordspacing}{\spaceskip=0pt\relax}
\providecommand{\BIBentryALTinterwordstretchfactor}{4}
\providecommand{\BIBentryALTinterwordspacing}{\spaceskip=\fontdimen2\font plus
\BIBentryALTinterwordstretchfactor\fontdimen3\font minus
  \fontdimen4\font\relax}
\providecommand{\BIBforeignlanguage}[2]{{%
\expandafter\ifx\csname l@#1\endcsname\relax
\typeout{** WARNING: IEEEtran.bst: No hyphenation pattern has been}%
\typeout{** loaded for the language `#1'. Using the pattern for}%
\typeout{** the default language instead.}%
\else
\language=\csname l@#1\endcsname
\fi
#2}}
\providecommand{\BIBdecl}{\relax}
\BIBdecl

\bibitem{hamadi2022hybrid}
R.~Hamadi~\textit{et al.}, ``A hybrid artificial neural network for task
  offloading in mobile edge computing,'' in \emph{IEEE International Midwest
  Symposium on Circuits and Systems}, 2022.

\bibitem{chen2022context}
B.~Chen, Z.~Yan, and K.~Nahrstedt, ``Context-aware image compression
  optimization for visual analytics offloading,'' in \emph{Proceedings of the
  13th ACM Multimedia Systems Conference}, 2022.

\bibitem{sun2022elasticedge}
H.~Sun~\textit{et al.}, ``{ElasticEdge: An Intelligent Elastic Edge Framework
  for Live Video Analytics},'' \emph{IEEE Internet of Things Journal}, 2022.

\bibitem{khan2022distributed}
M.~A. Khan~\textit{et al.}, ``Distributed inference in resource-constrained iot
  for real-time video surveillance,'' \emph{IEEE Systems Journal}, 2022.

\bibitem{yi2017lavea}
S.~Yi~\textit{et al.}, ``Lavea: Latency-aware video analytics on edge computing
  platform,'' in \emph{IEEE Symposium on Edge Computing}, 2017.

\bibitem{sun2022edgeeye}
H.~Sun~\textit{et al.}, ``{EdgeEye: A Data-driven Approach for Optimal
  Deployment of Edge Video Analytics},'' \emph{IEEE Internet of Things
  Journal}, 2022.

\bibitem{liu2018edge}
Q.~Liu~\textit{et al.}, ``An edge network orchestrator for mobile augmented
  reality,'' in \emph{IEEE Conference on Computer Communications}, 2018.

\bibitem{wang2020joint}
C.~Wang~\textit{et al.}, ``Joint configuration adaptation and bandwidth
  allocation for edge-based real-time video analytics,'' in \emph{IEEE
  Conference on Computer Communications}, 2020.

\bibitem{zhao2022edgeadaptor}
K.~Zhao~\textit{et al.}, ``Edge{A}daptor: Online configuration adaption, model
  selection and resource provisioning for {E}dge {DNN} inference serving at
  scale,'' \emph{IEEE Transactions on Mobile Computing}, pp. 1--16, 2022.

\bibitem{rost2017network}
P.~Rost~\textit{et al.}, ``Network slicing to enable scalability and
  flexibility in {5G} mobile networks,'' \emph{IEEE Communications magazine},
  2017.

\bibitem{ibm2021frameresolution}
\BIBentryALTinterwordspacing
IBM, \emph{{Camera frame rate, resolution, and video format requirements}},
  2021. [Online]. Available:
  \url{https://www.ibm.com/docs/en/video-analytics/1.0.6}
\BIBentrySTDinterwordspacing

\bibitem{andrews2011tractable}
J.~G. Andrews, F.~Baccelli, and R.~K. Ganti, ``A tractable approach to coverage
  and rate in cellular networks,'' \emph{IEEE Transactions on communications},
  vol.~59, no.~11, pp. 3122--3134, 2011.

\bibitem{peris2022modelling}
J.~Anguera-Peris \emph{et~al.}, ``{Modelling Multi-Cell Edge Video
  Analytics},'' in \emph{IEEE International Conference on Communications},
  2022.

\bibitem{glenn2021yolov5}
\BIBentryALTinterwordspacing
J.~Redmon and G.~Jocher, \emph{YOLOv5}, 2021. [Online]. Available:
  \url{https://github.com/ultralytics/yolov5}
\BIBentrySTDinterwordspacing

\bibitem{bochkovskiy2020yolov4}
A.~Bochkovskiy~\textit{et al.}, ``Yolov4: Optimal speed and accuracy of object
  detection,'' \emph{arXiv preprint arXiv:2004.10934}, 2020.

\bibitem{ameur2021deployability}
A.~B. Ameur, A.~Araldo, and F.~Bronzino, ``On the deployability of augmented
  reality using embedded edge devices,'' in \emph{IEEE 18th Annual Consumer
  Communications \& Networking Conference}, 2021.

\bibitem{baccelli2010stochastic}
F.~Baccelli and B.~B{\l}aszczyszyn, ``{Stochastic geometry and wireless
  networks: Volume II applications},'' \emph{Foundations and Trends in
  Networking}, vol.~4, no. 1--2, pp. 1--312, 2010.

\bibitem{iversen2010teletraffic}
V.~B. Iversen, ``Teletraffic engineering and network planning,''
  \emph{Technical University of Denmark}, p. 270, 2010.

\bibitem{tijms2003first}
H.~C. Tijms, \emph{A first course in stochastic models}.\hskip 1em plus 0.5em
  minus 0.4em\relax John Wiley and sons, 2003.

\bibitem{marler2010weighted}
R.~T. Marler and J.~S. Arora, ``The weighted sum method for multi-objective
  optimization: new insights,'' \emph{Structural and multidisciplinary
  optimization}, vol.~41, no.~6, pp. 853--862, 2010.

\bibitem{barry2000analytical}
D.~Barry~\textit{et al.}, ``{Analytical approximations for real values of the
  Lambert W-function},'' \emph{Mathematics and Computers in Simulation},
  vol.~53, no. 1-2, pp. 95--103, 2000.

\bibitem{abramowitz1988handbook}
M.~Abramowitz, I.~A. Stegun, and R.~H. Romer, ``Handbook of mathematical
  functions with formulas, graphs, and mathematical tables,'' 1988.

\bibitem{ghosh2010fundamentals}
A.~Ghosh~\textit{et al.}, \emph{Fundamentals of LTE}.\hskip 1em plus 0.5em
  minus 0.4em\relax Pearson Education, 2010.

\bibitem{ran2018deepdecision}
X.~Ran \emph{et~al.}, ``Deepdecision: A mobile deep learning framework for edge
  video analytics,'' in \emph{IEEE conference on computer communications},
  2018.

\end{thebibliography}
\bibliographystyle{IEEEtran}

\end{document}